\documentclass[10pt,conference]{IEEEtran}
\usepackage{cite}
\usepackage{amsmath,amssymb,amsfonts}

\usepackage{algpseudocode}
\usepackage{graphicx}
\usepackage{textcomp}
\usepackage{xcolor}
\usepackage[hyphens]{url}
\usepackage{fancyhdr}
\usepackage[breaklinks]{hyperref}
\usepackage{xspace}
\usepackage{array}
\usepackage{comment}
\usepackage{booktabs}
\usepackage{tikz}
\usepackage{siunitx}
\usepackage{multirow}

\usepackage[capitalize]{cleveref} 

\usepackage[linesnumbered,lined,vlined,boxed,commentsnumbered,ruled,longend]{algorithm2e}
\SetKwRepeat{Do}{do}{until}
\Crefname{algocf}{Algorithm}{Algorithms}

\SetNlSty{bfseries}{\color{black}}{}

\definecolor{olivegreen}{rgb}{0, 0.6, 0}
\definecolor{redorange}{HTML}{FF5349}
\definecolor{blue(ncs)}{rgb}{0.0, 0.53, 0.74}
\definecolor{navy}{HTML}{273BE2}

\usepackage{tcolorbox}

\newcommand{\hpcayear}{2025}
\newcommand{\thiswork}{Piccolo\xspace}
\newcommand{\ourcache}{\thiswork-cache\xspace}
\newcommand{\Ourcache}{\thiswork-cache\xspace}
\newcommand{\OurCache}{\thiswork-Cache\xspace}
\newcommand{\CQ}{Collection-Extended MSHR\xspace}
\newcommand{\Cq}{Collection-extended MSHR\xspace}
\newcommand{\cq}{collection-extended MSHR\xspace}

\newcommand{\rev}[1]{{\color{olivegreen}#1}}

\usepackage{marginnote}

\newcommand*\circled[1]{\tikz[baseline=(char.base)]{
            \node[shape=circle,draw,inner sep=0.4pt] (char) {#1};}}

\newcommand*\bbcircled[1]{\textbf{\textcolor{white}{\tikz[baseline=(char.base)]{
	    \node[shape=circle,draw,inner sep=0.4pt,fill=black] (char) {#1};}}}}

\definecolor{skyblue}{RGB}{195, 219, 237}
\newcommand*\bluecircled[1]{\tikz[baseline=(char.base)]{
        \node[shape=circle,draw=black,fill=skyblue,inner sep=1pt] (char) {\textcolor{black}{#1}};}}

\newcommand{\hpcasubmissionnumber}{35}
\title{\thiswork: Large-Scale Graph Processing with Fine-Grained In-Memory Scatter-Gather}

\def\hpcacameraready{} 

\newcommand\hpcaauthors{Changmin Shin\IEEEauthorrefmark{2}, Jaeyong Song\IEEEauthorrefmark{2}, Hongsun Jang\IEEEauthorrefmark{2}, Dogeun Kim\IEEEauthorrefmark{2}, Jun Sung\IEEEauthorrefmark{2}, Taehee Kwon\IEEEauthorrefmark{2}, Jae Hyung Ju\IEEEauthorrefmark{2}, \\ Frank Liu\IEEEauthorrefmark{3}, Yeonkyu Choi\IEEEauthorrefmark{4}, and Jinho Lee\IEEEauthorrefmark{2}
}
\newcommand\hpcaaffiliation{
\IEEEauthorrefmark{2}\textit{Department of Electrical and Computer Engineering, Seoul National University} \\
\IEEEauthorrefmark{3}\textit{School of Data Science, Old Dominion Univeristy} \\ 
\IEEEauthorrefmark{4}\textit{Samsung Electronics}

}
\newcommand\hpcaemail{$\{$scm8432, jaeyong.song, hongsun.jang, kdg6245, junsung3737, jessica314, hpotato$\}$@snu.ac.kr \\ fliu@odu.edu,  yeonkyuchoi7@gmail.com, 
leejinho@snu.ac.kr
}

\author{
  \ifdefined\hpcacameraready
    \IEEEauthorblockN{\hpcaauthors{}}
      \IEEEauthorblockA{
        \hpcaaffiliation{} \\
        \hpcaemail{}
      }
  \else
    \IEEEauthorblockN{\normalsize{HPCA \hpcayear{} Submission
      \textbf{\#\hpcasubmissionnumber{}}} \\
      \IEEEauthorblockA{
        Confidential Draft \\
        Do NOT Distribute!!
      }
    }
  \fi 
}

\fancypagestyle{camerareadyfirstpage}{%
  \fancyhead{}
  
  \fancyhead[C]{
    \ifdefined\aeopen
    \parbox[][12mm][t]{13.5cm}{\hpcayear{} IEEE International Symposium on High-Performance Computer Architecture (HPCA)}    
    \else
      \ifdefined\aereviewed
      \parbox[][12mm][t]{13.5cm}{\hpcayear{} IEEE International Symposium on High-Performance Computer Architecture (HPCA)}
      \else
      \ifdefined\aereproduced
      \parbox[][12mm][t]{13.5cm}{\hpcayear{} IEEE International Symposium on High-Performance Computer Architecture (HPCA)}
      \else
      \parbox[][0mm][t]{13.5cm}{\hpcayear{} IEEE International Symposium on High-Performance Computer Architecture (HPCA)}
    \fi 
    \fi 
    \fi 
    \ifdefined\aeopen 
      \includegraphics[width=12mm,height=12mm]{ae-badges/open-research-objects.pdf}
    \fi 
    \ifdefined\aereviewed
      \includegraphics[width=12mm,height=12mm]{ae-badges/research-objects-reviewed.pdf}
    \fi 
    \ifdefined\aereproduced
      \includegraphics[width=12mm,height=12mm]{ae-badges/results-reproduced.pdf}
    \fi
  }
  \fancyfoot[C]{}
}
\begin{document}
\maketitle

\ifdefined\hpcacameraready 
  \thispagestyle{camerareadyfirstpage}
  \pagestyle{empty}
\else
  \thispagestyle{plain}
  \pagestyle{plain}
\fi

\newcommand{\hpcaheight}{0mm}
\ifdefined\eaopen
\renewcommand{\hpcaheight}{12mm}
\fi

\newcommand\freefootnote[1]{%
  \let\thefootnote\relax%
  \footnotetext[0]{#1}%
  \let\thefootnote\svthefootnote%
}



\begin{abstract}
Graph processing requires irregular, fine-grained random access patterns incompatible with contemporary off-chip memory architecture, leading to inefficient data access.
This inefficiency makes graph processing an extremely memory-bound application.
Because of this, existing graph processing accelerators typically employ a graph tiling-based or processing-in-memory (PIM) approach to relieve the memory bottleneck. 
In the tiling-based approach, a graph is split into chunks that fit within the on-chip cache to maximize data reuse.
In the PIM approach, arithmetic units are placed within memory to perform operations such as reduction or atomic addition.
However, both approaches have several limitations, especially when implemented on current memory standards (i.e., DDR).
Because the access granularity provided by DDR is much larger than that of the graph vertex property data, much of the bandwidth and cache capacity are wasted.
PIM is meant to alleviate such issues, but it is difficult to use in conjunction with the tiling-based approach, resulting in a significant disadvantage. 
Furthermore, placing arithmetic units inside a memory chip is expensive, thereby supporting multiple types of operation is thought to be impractical.
To address the above limitations, we present \emph{\thiswork}, an end-to-end efficient graph processing accelerator with fine-grained in-memory random scatter-gather.
Instead of placing expensive arithmetic units in off-chip memory, \thiswork focuses on reducing the off-chip traffic with non-arithmetic function-in-memory of random scatter-gather.
To fully benefit from in-memory scatter-gather, \thiswork redesigns the cache and miss-handling architecture (MHA) of the accelerator such that it can enjoy both the advantage of tiling and in-memory operations.
\thiswork achieves a maximum speedup of 3.28$\times$ and a geometric mean speedup of 1.62$\times$, along with up to 59.7\% reduction in energy consumption across various and extensive benchmarks.

\end{abstract}


\section{Introduction}
\label{sec:intro}

\begin{figure}[t]
    \centering
    \includegraphics[width=\columnwidth]{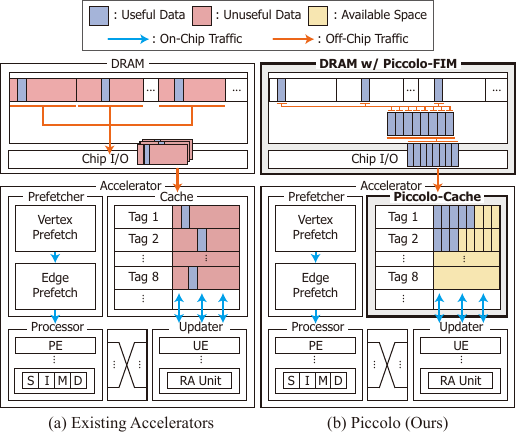}
    \caption{\thiswork overview compared to existing graph processing accelerators.}
    \vspace{-3mm}
    \label{fig:overall} 
\end{figure}

Graphs excel in handling non-structured data, as seen in social networks~\cite{campbell2013social,tang2010graph,majeed2020graph} and bioinformatics~\cite{bio1,bio2,kim2019graph}, offering enhanced expressive power over structured data formats.
However, graph processing is challenging due to its irregular, fine-grained random accesses, which is not well-suited to current off-chip memory (e.g., DRAM) architectures~\cite{ddr3,ddr4,hbm}.
This results in inefficient data accesses and makes graph processing a memory-bound application~\cite{graphicionado,scalagraph,random1}.

To address this, numerous graph processing accelerators~\cite{ozdal, graphicionado, graphdyns, scalagraph, graphpim, tesseract, graphq, graphp,mosaic,flashgraph} have been developed, roughly categorized into \emph{graph tiling-based accelerators} and \emph{processing-in-memory (PIM)}. 
In graph tiling-based accelerators~\cite{graphicionado,graphdyns,fabgraph,foregraph,stopcrying,graphit,largescaleFPGA,fabgraph2,fpgp}, as illustrated in \cref{fig:overall}a, the vertex data are partitioned into tiles that fit into on-chip caches.
This enhances data reuse at the cost of increasing raw access.
On the other hand, PIMs
~\cite{tesseract,graphpim,graphq,graphp} 
perform aggregation of vertices or atomic operations in memory to reduce the traffic. 

Both approaches have their own limitations due to their methodological characteristics.
When graph tiling~\cite{tiling} with on-chip memory is applied to maximize hits, they inadvertently retrieve unnecessary data from off-chip memory, which can be attributed to the discrepancy between off-chip burst length (64B for DDR family) and the access granularity (4B/8B vertex data), which is depicted in \cref{fig:overall}a with red (unuseful data).
Furthermore, the increased cache hit rates are not free, as they require more \emph{repetitions over tiles} that increase the number of raw memory accesses for redundant data load (e.g., graph topologies).
PIMs~\cite{graphpim, graphq, tesseract, graphp} appear to improve performance by exploiting the ample internal memory bandwidth inherent to PIM. 
However, they add a large area overhead to the memory die~\cite{hbmpim, aim, upmem_benchmark} and tend to suffer from being unable to utilize locality effectively.

In addition, it is widely recognized that simultaneously applying both approaches to get the benefits of each is non-trivial.
While PIM typically benefits from random accesses by computing inside the off-chip memory, graph tiling is a strategy employed to minimize these random accesses by using an on-chip cache.
The need for coherency between cache and memory further complicates the logic~\cite{pei, lazypim}.
Moreover, a majority of PIMs suffer from large arithmetic unit areas within a memory technology node.
Evidence from industry products indicates that supporting a single type (fp16) contributes approximately 50\% area overhead to the entire die~\cite{aim, hbmpim,gddraim}.
Such overheads are already challenging to accommodate for memory products oriented toward density, rendering the support for additional datatypes nearly unfeasible for
standard off-chip memories such as DDR~\cite{ddr3, ddr4}.

Here, we aim to address the above challenge of utilizing the internal bandwidth of off-chip memory at a low cost while benefiting from on-chip cache.
To accomplish this, we introduce \emph{\thiswork} (inspired by the concept of `pick and collect'), a fast graph processing accelerator that utilizes in-memory random scatter-gather.
As shown in \cref{fig:overall}b, the proposed \thiswork allows for fine-grained scatter-gather, where only the useful data are transferred and stored on caches.

First, we propose \emph{\thiswork function-in-memory (\thiswork-FIM)}. 
As previous PIMs have been hindered by the cost of arithmetic units and the challenge of integrating on-chip cache, we have opted to place \emph{no arithmetic units} in the off-chip DRAM.
Instead, \thiswork 
facilitates in-memory random scatter-gather to tackle the issue of random memory access by leveraging the abundant DRAM internal bandwidth.
The idea is to perform scatter/gather within banks, whose region is confined to a row. 
Because of this, the latency remains deterministic, and the data can be transferred through a single burst. 
While there have been several seminal works on supporting in-DRAM scatter/gather~\cite{gsdram,sam} or in-DRAM caching~\cite{figaro, tiered}, \thiswork differs in that it supports irregular accesses in a more fine-grained manner.
In addition, \thiswork is lightweight and is fully compatible with existing standard protocols. 

Second, we integrate \thiswork-FIM with an accelerator with on-chip cache via \emph{\ourcache}. 
Specifically, we devise a cache architecture that stores data in finer granularity than 64B lines, with an extension of MSHR to collect multiple requests into a single FIM operation. 
Many graph processing accelerators proposed in the literature~\cite{graphicionado, graphdyns, largescaleFPGA} rely on tiling~\cite{gridgraph} to enhance cache locality at the cost of repeated accesses.
With \thiswork, the fine-grained random access enables using much larger and sparser tiles.
Combined with the fact that \thiswork does not offload any computation to PIM, \thiswork seamlessly benefits from both the PIM and cache locality. 

To demonstrate \thiswork's effectiveness, we benchmarked it against a wide range of tiling-based graph processing accelerators and PIMs.
The results show that \thiswork can achieve up to a 3.28$\times$ speedup and reduce energy consumption by up to 59.7\% compared to the baseline graph accelerator~\cite{graphdyns} with a conventional system. Additionally, we emulated \thiswork on an FPGA platform to verify the compatibility of \thiswork commands with standard DRAM commands.

The contributions can be summarized as follows:
\begin{enumerate}
    \item We introduce \thiswork-FIM, in-memory random scatter-gather, requiring no arithmetic units on off-chip memory, to utilize internal memory bandwidth at a low cost.
    \item We integrate \thiswork-FIM with on-chip cache using \ourcache, gaining advantages from \thiswork-FIM without the need for a user program modification. 
    \item We validate the compatibility of \thiswork commands with standard DRAM commands through an FPGA emulation.
    \item \thiswork provides up to a 3.28$\times$ speedup and 59.7\% energy reduction compared to the prior arts.    
\end{enumerate}

\section{Background}

\subsection{Graph Processing Model}
\label{sec:vcm}

For various purposes, graph processing~\cite{pagerank} is often described using diverse variations of processing models~\cite{pregel,graphmat,powergraph,tao,ligra,sisa,graphblas} due to its ease of programming, improved performance, and efficient scalability. 
Among them, the vertex-centric model (VCM)~\cite{pregel} is the most widely used one for parallel graph processing~\cite{tesseract, localitygraph, graphpim, graphp, graphq, scalagraph}.

\SetInd{0.6em}{0.7em}
\setlength{\algomargin}{8pt}

\begin{algorithm}[t]
    \caption{Graph Processing Iteration with Tiling}
    \label{algo:vcpm}
    \DontPrintSemicolon
    \SetNoFillComment
    \SetKwBlock{DoParallel}{do in parallel}{end}
    \SetKwInOut{Input}{Input}
    \SetKwInOut{Output}{Output}
    \SetKwProg{Fn}{Function}{}{end}
    \vspace{.5mm}
    \Input{\small{$G = (V, E)$ - Input Graph\\
            $V_{prop}$ - Vertex Property Array\\
            $V_{active} \subset V$ - Active Vertex Set\\
            $V_{active}' = \emptyset$ - Active Vertex Set of Next Iteration}}
    \vspace{1mm}
    \Output{\small{$V_{active}'$ - Active Vertex Set of Next Iteration}}
    \vspace{2mm}
    
    \ForEach(\tcp*[f]{\small Tiling (Optional)}){\upshape $V_{tile} \subset$ \upshape $V$ }  
    {
    \ForEach{\textbf{\upshape $u$} $\in$ \textbf{\upshape $V_{active}$}}
    {
        \ForEach{$e=(u,v)\in E, v \in V_{tile}$}
        {
            $res$ = \textbf{Process}($e$.weight, $V_{prop}$[$u$]) \\
            $V_{temp}$[$v$] = \textbf{Reduce}($V_{temp}$[$v$], $res$)\\
              
        }
    }

    \ForEach{$v$ $\in$ $V_{tile}$}
    {
        $applyres$ = \textbf{Apply}($V_{prop}$[$v$], $V_{temp}$[$v$], $V_{const}$[$v$]) \\
        \If{$V_{prop}$ \textnormal{[$v$] !=} $applyres$}
        {
            $V_{prop}$[$v$] = $applyres$ \\
            \upshape $V_{active}'$ = \upshape $V_{active}' \cup v$ \\
        }
         
    }

    }
    
\end{algorithm}

\begin{figure}[t]
    \centering
    \includegraphics[width=\columnwidth]{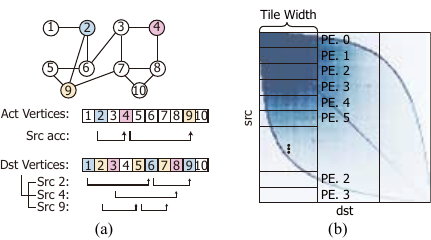}
    \caption{(a) Random accesses pattern and (b) tiling method in graph processing.}
    \label{fig:background} 
\end{figure}

\cref{algo:vcpm} illustrates an example iteration of graph processing in VCM utilizing three essential operators: \emph{Process}, \emph{Reduce}, and \emph{Apply}. 
These are application-defined functions that differ depending on graph algorithms.
During each iteration, each active vertex is visited, and all its edges are traversed (lines 2-3).
With the edge weight and vertex property ($V_{prop}$), each temporary vertex property ($V_{temp}$) is updated by $process$ and $reduce$ (lines 4-5).
After traversing all the edges, each vertex updates its $V_{prop}$ using $V_{temp}$ and, if needed, a constant value $V_{const}$ (lines 6-7).
Finally, if the $V_{prop}$ is changed, the vertex is activated for the next iteration (lines 8-10).
Meanwhile, due to the random nature of graphs, traversing edges exhibit irregular random memory access to the $V_{temp}$ array.
For example, \cref{fig:background}a shows the example of memory access patterns in graph processing.
When traversing the active vertex \bluecircled{2}, the neighbors of \bluecircled{2} (\circled{1}, \circled{6}, \circled{9}) are visited, and those memory accesses are completely random.
The other active vertices exhibit the same random access patterns.
Moreover, the amount of computation on randomly accessed data is small, requiring high memory bandwidth.


\subsection{Challenges in Graph Processing}

To mitigate the memory access bottleneck of graph processing, existing approaches primarily employ aggressive prefetching to hide memory latency~\cite{bingo}.
\cref{fig:overall}a shows a typical graph processing architecture ~\cite{ozdal, foregraph, graphicionado, graphdyns, graphpulse, scalagraph}, consisting of a prefetcher, a processor, and an updater with on-chip memory. 
From \cref{algo:vcpm}, the prefetcher continuously loads the graph topology and the corresponding vertex properties ($V_{prop}$ and $V_{temp}$) from edges $e$ in line 2-3.
Then, the processor operates the $process$ function shown in line 4 with the prefetched data.
After passing the crossbar switch for parallel atomic updates, the updater executes $reduce$ with prefetched $V_{temp}[u]$ (line 5).

With sufficient prefetching to hide latencies, the bottleneck moves to the memory bandwidth cost from three components: topology read, sequential property access, and random property access (e.g., $V_{active}=V$). 
The topology consists of accessing the CSR format, which is proportional to $|V|$ for row indices and $|E|$ for column indices.
Then, assuming the edges $(u,v)$ are ordered by its source $u$ (i.e., push approach), $V_{prop}[u]$ is sequentially accessed, and $V_{temp}[v]$ is randomly accessed. 
The sequential access cost is proportional to $|V|$, and the random access cost is proportional to $|E|$ times the burst size, assuming that the data are larger than the cache.

From these, graph tiling~\cite{mosaic,flashgraph,gridgraph} is a popular approach to reduce the random access cost. 
As illustrated in \cref{fig:background}b, restricting the destination vertices to a certain range can enhance locality.  
When the tile width is smaller than the cache capacity (hereafter called \emph{perfect tiling}), the random access cost dramatically reduces to be proportional to $|V|$.
However, this comes at the cost of increased repetition on topology and sequential accesses. 
As the source vertex $u$ is accessed once per tile, its cost with $t$ tiles increases to be proportional to $t|V|$. 
Furthermore, the row indices separately exist for each tile, increasing the row index cost again by $t$ times.

Because of this, there usually exists a sweet spot that finds the balance between locality and repetition.
With \thiswork, the cost of random accesses within a tile would be greatly reduced because of the fine-grained accessing and caching. 
Moreover, this contributes to moving the sweet spot to have larger tiles (hence smaller $t$), achieving additional speedup.

\subsection{DRAM Architecture and Timing Parameters}
A simplified diagram of the modern DRAM hierarchy is shown in \cref{fig:dram} by the unshaded boxes. 
The host can utilize one or more DRAM channels (\emph{Host\&Bus}). 
Each channel has a dedicated command, address, and data bus.
One or more memory chips can be connected to each DRAM channel.
An example organization in \cref{fig:dram} includes four chips, each equipped with x16 pins.
Given that the data output width of each DRAM chip is 16 bits, multiple chips are grouped together to form a rank.
All chips within a rank share the command and address buses, but each chip has its own dedicated data bus.
Consequently, any command sent to a rank is processed by all the chips within the rank to provide a 64-bit data width.
Each chip contains multiple banks (Banks 0-7) arranged in an array, with each bank consisting of numerous rows that hold multiple cache lines identified by columns.

To access the data in DRAM, a row from the data cell array is first activated and transferred to the sense amplifiers (`Activate', \emph{ACT}). 
The latency between the start of activation and the availability of data is $tRCD$. 
Second, to access a cache line from the activated row, the memory controller issues (`Read', \emph{RD}) or (`Write', \emph{WR}) with the column address.
The distance between two consecutive RD/WR for the row is $tCCD$ or $tBURST$.
When $tRAS$ after activation or $tWR$ after writing data burst, the memory controller can precharge the bank to activate a different row (`Precharge', \emph{PRE}).
The precharge for the next activation takes $tRP$.





\section{Motivational Study}
\label{seC:moti}

\begin{figure}[t]
    \centering
    \includegraphics[width=\columnwidth]{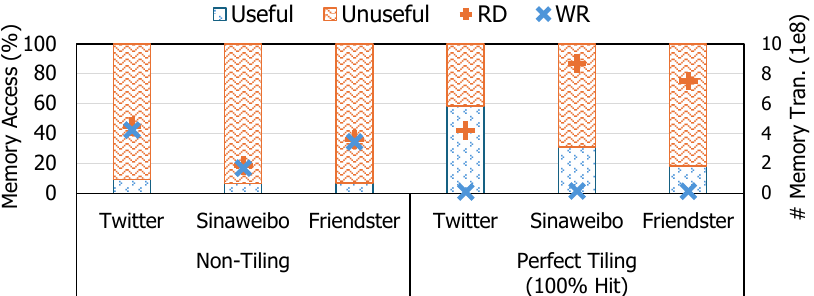}
    \caption{Motivational experiment on BFS algorithm. Existing accelerators still suffer from unnecessary accesses due to fine-grained random access, even with perfect tiling, which brings full cache hits.} 
    \label{fig:moti} 
\end{figure}

%
\cref{fig:moti} depicts a motivational experiment designed to emphasize the necessity of a holistic method on top of current tiling-based accelerators.
It illustrates the limitations of current tiling-based techniques in utilizing the on-chip cache.
We show the memory access of a graph accelerator~\cite{graphdyns} running 
Breath-First Search (BFS) algorithm with non-tiling and perfect tiling, which makes 100\% cache hit except for cold misses.
The breakdown is represented through a bar graph (left axis) that differentiates between useful and unuseful memory access.
In addition, we depict the number of read (RD) and write (WR) transactions using dots (right axis).
For a comprehensive understanding of the experimental setup, please refer to \ref{sec:method}.

As depicted in the figure, non-tiling methods suffer from highly randomized accesses to the vertex data.
Because the memory access granularity is much smaller than the individual vertex data ($64B$), over 90\% of the accessed data are evicted from the cache without being used. 
This not only results in severe cache pollution but also wastes precious memory bandwidth.
Furthermore, since the BFS algorithm accesses only a subset of graph vertices (active vertices) in each iteration, the sparsity of the algorithm exacerbates the inefficient use of cache capacity.
To alleviate this, many accelerators~\cite{graphicionado, graphdyns, scalagraph, largescaleFPGA} adopt perfect tiling that confines the working set into a single tile.
This indeed improves the locality, as shown in the right part of the figure. 
However, it comes at the cost of the highly increased number of read accesses due to the inherent topology read repetition of the tiling approach.

This indicates that there is great room for speedup by supporting fine-grained access from the off-chip memory.
It would make more efficient use of the memory bandwidth, and the cache would be able to capture the locality better.
Unfortunately, the existing memory subsystems are highly optimized for coarse-grained accesses to provide better bandwidth and latency for general cases that tend to exhibit more sequential accesses.
To address such issues, we propose designs for both function-in-memory (\cref{sec:main}) and cache architecture (\cref{sec:cache}) with thorough evaluations (\cref{sec:eval}).

\section{\thiswork Function-in-Memory: In-DRAM Random Scatter-Gather}
\label{sec:main}

\subsection{Overview}
For graph processing, supporting fine-grained (e.g., 8B) accesses from DRAM would greatly enhance the performance by reading and writing random vertex data. 
However, the current memory subsystem does not allow such accesses, as their design is oriented around a fixed-length \emph{burst}, which spans 64 bytes for DDR standards.
Naive attempts to modify the current DRAM architecture with fine-grained read/write operations would result in a significant overhead on the command bus with little benefit from data bandwidth.

Thus, the main idea behind \thiswork-FIM has two aspects:
sending offsets to gather and scatter through the data bus, and secondly restricting their operation to a single row.
Instead of sending all the addresses over the command bus, we treat them as data and transfer them over the data bus to a special offset buffer with a simple write command.
We further restrict the area of a single scatter/gather within a single row of a bank. 
This has several benefits.
First, it helps achieve high bandwidth. 
Because row activation is one of the most expensive DRAM operations, the DRAM is designed to hide such latency.
By restricting the range of a scatter/gather operation only within a row, we can ensure the fine-grained read/write operation is not disturbed by unnecessary activations for other rows, thus achieving high performance.
Secondly, it aids in achieving deterministic latency.
If activations are needed during the operation, it will be extremely difficult for the memory controller to determine the number of activations needed. 
The resulting long wait time becomes another difficult timing parameter for the memory controller.

\subsection{Microarchitecture and Procedure}

\begin{figure}[t]
    \centering
    \includegraphics[width=\columnwidth]{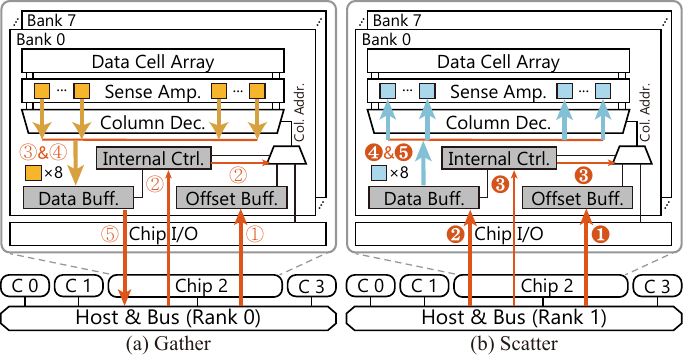}
    \caption{\thiswork architecture for (a) gather and (b) scatter operations. Shaded boxes depict the newly added modules on top of conventional DRAM.}
    \label{fig:dram} 
\end{figure}

\cref{fig:dram} shows the architecture of \thiswork on top of modern DRAM, where the new components are shown as shaded gray. 
\thiswork-FIM adds three simple components inside each DRAM: an offset buffer, a data buffer, and an internal controller.
For convenience, we introduce five new commands, namely \textsc{write offset buffer}, \textsc{gather execute}, \textsc{scatter execute}, \textsc{read data buffer}, and \textsc{write data buffer}.
We provide a detailed description of the execution of these commands without the need to modify the DDR protocol in \cref{sec:main:impl}.

\cref{fig:dram}a shows how eight 64-bit items are gathered using a \thiswork-FIM operation on $\times$16 DDR4 devices.
\circled{1} The host sends the eight offsets of the desired words within an already activated row through the write offset buffer command.
The offsets are sent over the data bus.
We use a 16-bit offset, which is sufficient to cover the entire row.
In DDR, a word is interleaved across multiple chips in units of the device width (16 bits for $\times$16 devices). 
This makes all devices access equal offsets, simplifying the operation. 
It also indicates that the offsets need to be duplicated across all chips, leading to a total of 128$n$ bits (16 bits per offset $\times$ 8 offsets $\times$ $n$ chips).
In the case of using $\times$16 DDR4 devices, a total of 512 bits (128 bits per chip $\times$ 4 chips) is needed, matching the data burst length.

Multiple bursts may be involved for devices with fewer pins (and hence more chips).
\circled{2} The host initiates the in-bank gather.
\circled{3} On the first offset, the internal controller issues a column read and picks the needed 64 bits (16 bits per chip $\times$ 4 chips) to the data buffer. 
\circled{4} The remaining seven offsets are repeated in the same manner as in the third step.
\circled{5} The host executes a data buffer read command, which sends the eight items gathered in the data buffer through the data bus to the host.
This procedure only consumes two data bus transfers: one to write offset buffers, and another to read data buffers. 
In a conventional DRAM, this takes eight normal reads, and thus \thiswork-FIM can achieve 4$\times$ bandwidth gain in the ideal case.
\cref{fig:dram}b shows the procedure for scattering eight 64-bit items.
\bbcircled{1} Similarly, the host writes the column offsets to the offset buffer.
Also, the host writes the data to the data buffer before initiating the scatter command (\bbcircled{2}), and there is no need for additional data buffer read afterward.
\bbcircled{3}-\bbcircled{5} Like the gather operation, the internal controller issues a column write and repeats for the eight columns.
The scatter operation also exhibits a theoretical bandwidth gain of 4$\times$ over the existing writes.


\section{\OurCache: Integration of \thiswork-FIM with On-Chip Memory}
\label{sec:cache}
In conjunction with in-memory random scatter-gather operations (\thiswork-FIM), we introduce \ourcache to fully utilize the gathered data. 
Conventional cache memory systems are incapable of storing the gathered data within a cache line because a cache line retrieved via \thiswork-FIM spans a non-contiguous address range. 

Furthermore, conventional caches can only issue burst-sized (e.g., 64B) memory requests.
In that case, the only option to utilize \thiswork would be to exploit the scatter-gather operation to a user program, incurring a non-negligible design overhead.
To address this limitation, we propose \ourcache, which stores and requests cache line data in a fine-grained granularity to take advantage of \thiswork-FIM.

\subsection{Cache Architecture}
\label{sec:cache_arch}

\begin{figure}[t]
    \centering
    \includegraphics[width=\columnwidth]{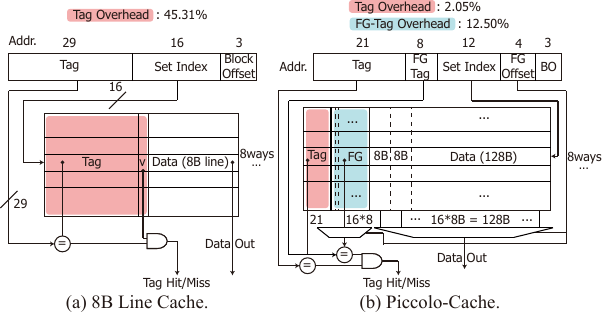}
    \vspace{-7mm}
    \caption{Implementations of 4MB and eight-way cache for 8B granularity data access with 48bit addressing.
     (a) 8B line cache. (b) \Ourcache. 
     While the 8B line cache suffers from the significant tag overhead, \thiswork addresses it with \ourcache.
     }
     \vspace{-3mm}
    \label{fig:cachearch} 
\end{figure}

To manage the 8B-granularity data, one can naively use a cache whose cache line size is 8B or a sectored cache~\cite{sectored, sectored2} comprised of 8B sectors. 
For example, \cref{fig:cachearch}a illustrates the structure of an 8B-line cache.
The 8B-line cache can manage the fine-grained data per cache line, and thus, the cache can hold only the useful data.
However, the 8B-line cache needs to store a tag for every single 8B data, which adds 8 times overhead to that of a conventional 64B-line cache.
For example, assuming an 8-way 4MB cache using 48-bit address space, the tag overhead can be calculated by 29 bits $\times$ 512K cache lines.
The total tag storage overhead of the 8B-line cache is nearly half of the total cache capacity. 
On the other hand, the sectored cache is composed of 64B cache lines, which is the same as the conventional cache.
If each cache line is composed of 8B sectors sharing the tag, 
cache data can be managed in smaller granularity only with the addition of one valid bit per 8B item.
However, the sectored cache needs to allocate an entire cache line even for a single sector, resulting in inefficient cache capacity usage.
This turns out to be detrimental to performance, as we will discuss in \cref{subsec:cachepolicy} in detail.

To address the above issues, we propose \ourcache illustrated in \cref{fig:cachearch}b.
We split a portion of the tag into a `fine-grained tag' (fg-tag) and associate it with each 8B sector.
The key observation behind this is that many cache lines contain cache line tags with low dynamic range thanks to the graph tiling approach.
By allocating a tag for a cache line and allocating a fg-tag in a sector granularity, we can mitigate the tag capacity overhead depending on the size of the fg-tag.

As shown in \cref{fig:cachearch}b, \ourcache first compares the address tag within the tag of the indexed cache line.
Second, the fg-tag and the sector data (8B) are selected by fine-grained offset bits (FG Offset).
Last, the address fg-tag is compared with the selected fg-tag in the cache, and it is determined whether the request is hit or miss.
By splitting the conventional tag into two regions (tag and fg-tag), we can reduce the tag storage overhead from being proportional to the full tag size to the smaller fg-tag size.
Moreover, unless the tag changes, \ourcache can operate as if 8B line cache because the data is indexed by set index (12 bits) and fg-offset (4 bits), which is the same as the set index (16 bits) of the 8B line cache.

To balance performance and tag overhead, we need to appropriately set the value of the number of fg-tag bits, and the size of a cache line.
To avoid too many evictions of the cache line instead of the fine-grained sector, we set the fg-tag bits as 8 bits.
Also, we set a single cache line to contain 16 sectors (a total of 128B per cache line) since the number of tags can be reduced proportional to the number of cache lines.
In the example design of \cref{fig:cachearch}b above, a 128B cache line stores 16 sectors of 8B data from a range of 32KB (15 address bits from fg-tag (8) + fg-offset (4) + byte offset (3)) that share the same tag.
To provide flexibility, we allow the same tags to appear multiple times within the same set (up to 8 times for an 8-way cache). 
This allows \ourcache to adapt to varying sparsities from diverse workloads and datasets.
To reduce the overhead of searching for fg-tags from potentially multiple ways with matching tags, we make the search sequential, where the ways are examined one by one. 
While this slightly increases the latency, its impact is almost negligible because of the throughput-oriented nature of graph processing.

\subsection{Fine-Grained Cache Replacement}
\label{subsec:cachepolicy}

\begin{figure}[t]
    \centering
    \includegraphics[width=\columnwidth]{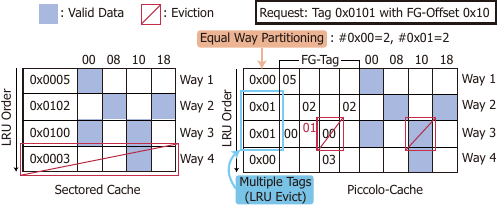}
    \caption{Example of cache line access and eviction of \ourcache compared to sectored cache.}
    \label{fig:cache_example} 
\end{figure}

In this subsection, we show how cache lines and sectors are replaced in \ourcache in comparison to a sectored cache.
\cref{fig:cache_example} demonstrates an example. 
For clarity, in our example, we simplified a cache into a four-way set associative cache, and each horizontal line refers to each way.
Starting from the initial state, a request of tag \texttt{0x0101} with FG-Offset \texttt{0x10} is received.
In the case of a sectored cache (left), the entire cache line of \texttt{0x0003} tag is evicted (red box) following the LRU order.
On the other hand, in \ourcache (right), the same tags (e.g., \texttt{0x01}) appear multiple times, as discussed in \cref{sec:cache_arch}.
\Ourcache conducts a sequential search for the tag \texttt{0x01} and then finds the target cache line (the third line in the example) following the LRU order among the lines of the same tags.
Then, the fg-tag (\texttt{0x00}) in the FG-offset (\texttt{0x10}) position does not match with the fg-tag of the request (\texttt{0x01}), leading to the eviction of that sector (red box).
Compared to the sectored cache case where a single sector occupies an evicted cache line, \ourcache evicts only a small single sector, thereby offering efficient cache capacity usages.

One remaining issue is determining when to evict the entire line to accommodate a differently tagged line. 
In a naive thought, a simple LRU seems to suffice.
However, because \thiswork-cache allows multiple lines with the same tag, sometimes a whole line replacement is needed even when a matching tag is found.
With no such consideration, any data covered by a single tag will occupy only up to one way of the cache. 
Thanks to graph tiling, we can pre-identify the list of tags that correspond to each tile range. 
From the identified range, we apply way partitioning to the tags within the tile. 
Thus, when a fg-tag miss occurs, if the corresponding tag does not occupy the allocated number of ways, an entire line occupied by another tag in LRU is evicted to install the new data.
While unequal partitioning~\cite{ucp} could be applied to improve the performance, we leave such policy as future work.


\subsection{\CQ}
\label{sec:cq}


To take advantage of \thiswork-FIM, issuing a collected request of read/write for the same DRAM row is crucial.
The challenge lies in the fact that such requests may come from multiple sets of the cache.
One might attempt to align the row addresses with the cache set addresses, but such a design would harshly increase cache conflicts and result in severe performance degradation.

Inspired by the state-of-the-art MSHR design~\cite{stopcrying} and victim caches,
we propose \emph{\cq}, an extended MSHR to generate collected requests from the host side.
The main idea behind \cq is to collect the misses from \thiswork-cache that belong to a single DRAM row such that they can be served by \thiswork-FIM.

\begin{figure}[t]
    \centering
    \includegraphics[width=\columnwidth]{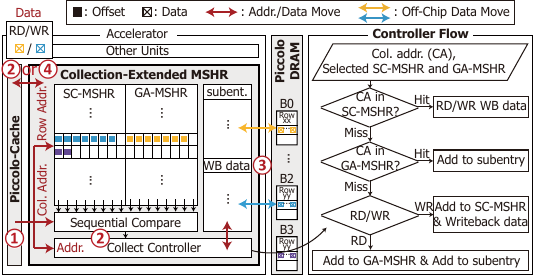}
    \vspace{-5mm}
    \caption{\Cq and its execution flow of the controller.}
    \vspace{-3mm}
    \label{fig:cbuff} 
\end{figure}

\cref{fig:cbuff} shows an example of our \cq design, which consists of a direct-mapped MSHR buffer with 16 column offset entries (half of the entries are for gather: GA-MSHR and the others are for scatter: SC-MSHR) and the direct-mapped cache for MSHR subentries and write-back data.
\circled{1} When the cache miss occurs, the miss request and possibly a write-back request are sent to the \cq.
Inside, the MSHR is indexed by the DRAM row address.
\circled{2} If the row address is found, the column offset of the request is compared with the existing offsets within the MSHR buffer.
If not, a buffer is newly allocated, possibly evicting another that invokes a partially filled gather or scatter.
We make the offset search sequential, and based on the matching results, the collect controller operates in the following order.
First, as shown in \cref{fig:cbuff}, if the incoming column offset hits in SC-MSHR, the request is served by the write-back data.
Second, if the incoming column offsets hit in GA-MSHR, this is equal to MSHR hit, thus just the subentry is stored inside the \cq.
Last, if there's no matching column offset, the request's column offset and either subentry or write-back data are stored inside \cq depending on whether the request is READ or WRITE.
\circled{3} If eight column offsets for either gather or scatter are collected, \cq executes the in-memory scatter/gather operation by using \thiswork-FIM.
\circled{4} The retrieved data for the read request is sent to the cache.

It is worth noting that the use of graph tiling \cite{gridgraph} prevents issuing too many \cq evictions. 
While \cq eviction typically does not become the system bottleneck, we find that carefully tuning the tile size to be moderate for what \cq can collectively cover helps gain slightly better performance. 
Note that \thiswork does not introduce any new cache coherency issue~\cite{coherency} because \thiswork-FIM does not change the value at the memory side.
The writeback data can be served from the buffer by the controller policy described in \cref{fig:cbuff} (right, Controller Flow).

\begin{figure}[t]
    \centering
    \includegraphics[width=\columnwidth]{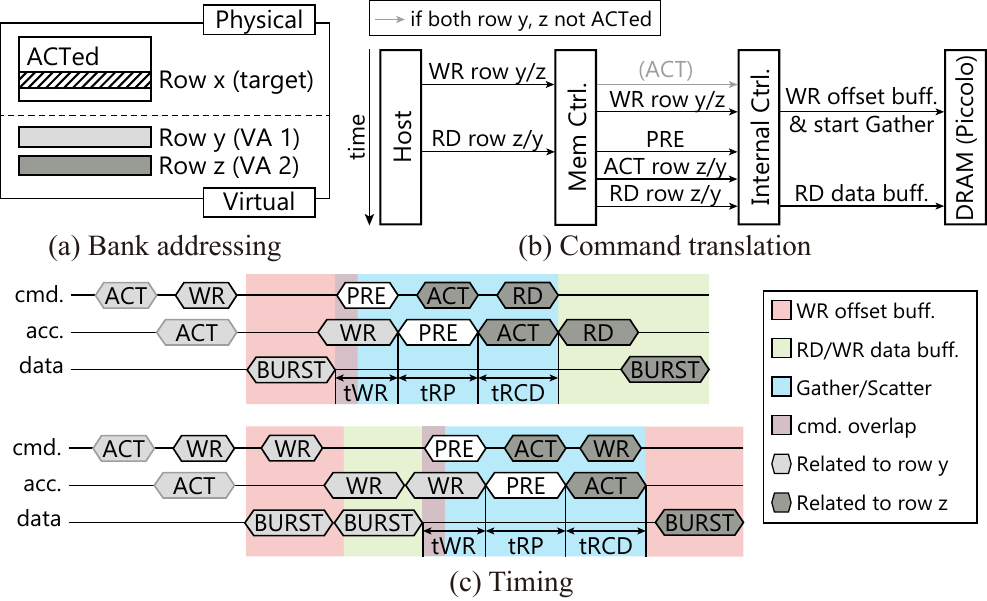}
    \caption{Operations of \thiswork-FIM. 
    In this figure, row $y$ corresponds to the offset buffer, and row $z$ corresponds to the data buffer (can be used interchangeably).
     (a) bank addressing. (b) command translation example for gather. (c) timing (scatter/gather).}
    \label{fig:cmd} 
\end{figure}

\section{Implementation: \thiswork Operations within DDR Standard}
\label{sec:main:impl}
To utilize \thiswork-FIM, the host memory controller should support the new operation added on DRAM.
\cref{fig:cmd} shows how the additional commands are implemented within the DDR standard.
In a naive way, we can add new commands, with one possible solution of utilizing the RFU (reserved for future use) opcodes in the standard~\cite{gradpim}.
However, adding new commands requires a significant change in the DDR protocol and the memory controller, which adds some design overhead to developing a custom memory controller. 
Instead, we \textit{only use the existing DRAM commands} to operate \thiswork-FIM.
The key idea is to assign virtual rows per bank and internally interpret normal reads/writes to them as special commands.

As described in \cref{fig:cmd}a, the address space of each bank is assigned to two virtual rows, depicted as $y$ and $z$.
A virtual row has two regions, which are mapped to the data buffer and offset buffer within the bank.
By assigning addresses to the buffers, reading/writing can be done via normal read/write operations.
However, the problem remains on the scatter/gather.
To access 8 words from the target row $x$ without any external data movement, the operation should occupy the bank for $8\times tCCD$ latency, which appears difficult to execute without adding a new command.
\thiswork-FIM addresses this by mapping the two virtual rows to the same pair of buffers and exploiting the activation latency between them to conceal the difference from the perspective of the memory controller.

\cref{fig:cmd}b shows how the commands for a gather operation are interpreted at each level with their timing depicted on \cref{fig:cmd}c. 
To perform a gather to an activated row $x$, the host performs a WR command to the offset buffer address in one of the virtual rows (e.g., $y$), whose range is exposed to the system software.
The data are written to the offset buffer (shaded red), and this automatically triggers the internal gather operation (shaded blue).
To read the gathered items from the data buffer, the host issues a read operation to the data buffer address of the other virtual row (e.g., $z$). 
Because row $z$ appears close to the memory controller, 
precharge and activation commands are sent for the row $z$. 
Since row $z$ is virtual, those commands are translated to a no-op by the internal controller.
Instead, this creates a $tWR+tRP+tRCD$ time gap for the internal gather operations.
Because $8\times tCCD$ equals around 39.84~\SI{}{\nano\second} and $tWR+tRP+tRCD$ is around 41.64~\SI{}{\nano\second} (DDR4-2400R~\cite{ddr4}),  there is sufficient time for the scatter/gather. 
For some products where $8\times tCCD$ is longer, we slightly adjust $tWR$. 
This introduces a small overhead in normal WR, which can be mostly hidden by bank parallelism.
Finally, the host receives the gathered data by an RD command (shaded green).

Likewise, to perform a scatter operation to an activated row $x$, the host performs a WR command to the data buffer after writing the offset buffer.
In this case, the host sends the WR commands for the internal buffers in the same virtual row (e.g., $y$).
Then, the next WR command to the offset buffer of the other virtual row (e.g., $z$), which is for another scatter/gather operation, causes the PRE and ACT.
This also creates a $tWR+tRP+tRCD$ time gap for internal scatter operations.
In cases where no command is scheduled for the internal buffer after the scatter operation, the memory controller sends a dummy write request to keep the activation delay.

\begin{table}[t]
    
    \centering 
    \caption{Evaluation Platforms}
    {
    \resizebox{0.9\columnwidth}{!}
    {
    \setlength{\tabcolsep}{3pt}
    \begin{tabular}{lcc}
        \toprule
     \textbf{Objective} & \textbf{Component} & \textbf{Evaluation Platform} \\  
    \midrule
     Validation (\S\ref{sec:emul}) & \thiswork-FIM & FPGA Emulation  \\  
     \cmidrule(lr){1-3}
     Performance (\S\ref{sec:eval:perf}$-$H) & Overall & Cycle-Accurate Simulator  \\ 
     \cmidrule(lr){1-3}
     \multirow{3}{*}{Energy and Area (\S\ref{sec:energy})} & Accelerator & RTL Synthesis \\ 
     & SRAM & CACTI 7.0~\cite{cacti} \\
     & DRAM & Design Comparison with~\cite{hynix} \\
    \bottomrule
    \end{tabular}
    } 
    } 
    \label{tab:evalmethod}
\end{table}

\section{Evaluation}
\label{sec:eval}
\subsection{Experimental Methodology}
\label{sec:method}
To evaluate \thiswork, we have performed functionality validation, performance measurement, and energy/area estimation. The methods are summarized in \cref{tab:evalmethod}.

\textbf{Validation and Feasibility Check.}
For validation and feasibility check of \thiswork-FIM with the DDR4 protocol, we conducted an FPGA emulation using a platform similar to PiDRAM~\cite{pidram} and PiMulator~\cite{pimulator}.
The emulation platform was constructed on an AMD ALVEO U280~\cite{u280} board.
On the FPGA, there is a memory controller following DDR4 standard~\cite{ddr4}. 
The memory controller is connected to 16 DRAM banks whose virtual row buffers are implemented using BRAMs. 
The bank data are stored within the HBM memory connected to the FPGA, which provides enough bandwidth and capacity to model bank-internal operations required for \thiswork-FIM.
We used $tCCD\_L$= 6nCK, $tCCD\_S$= 4nCK, $tRAS$= 39nCK, and $tBURST$= 4nCK for timing parameters.

\textbf{Performance.}
For measuring the performance of the baselines and \thiswork, we used an in-house cycle-accurate simulator for graph processing accelerator, briefly illustrated in \cref{fig:overall}.
There are eight PEs in total, each with 8-way SIMD lanes running at 1GHz.
We utilized four-rank DDR4-2400R x16 devices as the default for the memory system, which is simulated with Ramulator~\cite{ramulator}.
We set 4MB cache size as a default for the accelerator whose architecture mimics the baseline architecture of \cite{graphdyns}. 
We utilize collection-extended MSHR with 4K entries, following~\cite{largescaleFPGA}.

\begin{table}[t]
    \centering 
    \caption{Graph Datasets Used in the Evaluations}
    {
    \resizebox{.9\columnwidth}{!}
    {
    \setlength{\tabcolsep}{3pt}
    \begin{tabular}{lcccc}
        \toprule
     \textbf{Graph} & \textbf{$\#$Vertices} & \textbf{$\#$Edges} & \textbf{Brief Explanation} \\  
    \midrule
     Uci-Uni (UU) \cite{nr} & 58M & 92M    & Facebook Friendship\\
     \cmidrule(lr){0-3}
     Sinaweibo (SW) \cite{nr} & 21M  & 261M  &  Sina Weibo Social \\ 
     \cmidrule(lr){0-3}
     Twitter (TW) \cite{snap}   & 41M & 1465M  & Twitter Follower \\ 
     \cmidrule(lr){0-3}
     Friendster (FS) \cite{snap} & 65M & 1806M & Friendster Social\\ 
     \cmidrule(lr){0-3}
     Papers (PP) \cite{ogbn}  & 111M & 1615M & Citation \\
     \cmidrule(lr){0-3}
     Watts–Strogatz scale 26 (WS26) \cite{watzstrogatz}  & 67M & 336M & Synthetic Graph \\
     \cmidrule(lr){0-3}
     Watts–Strogatz scale 27 (WS27) 
     \cite{watzstrogatz}  & 134M & 671M & Synthetic Graph \\
     \cmidrule(lr){0-3}
     Kronecker scale 25 (KN25) \cite{kronecker}  & 34M & 336M & Synthetic Graph \\
     \cmidrule(lr){0-3}
     Kronecker scale 26 (KN26) \cite{kronecker}  & 67M & 671M & Synthetic Graph \\
     \cmidrule(lr){0-3}
     Kronecker scale 27 (KN27) \cite{kronecker}  & 134M & 1342M & Synthetic Graph \\
     \cmidrule(lr){0-3}
     Kronecker scale 28 (KN28) \cite{kronecker}  & 268M & 2684M & Synthetic Graph \\
     
    \bottomrule
    \end{tabular}
    } 
    }
    \label{tab:dataset}

\end{table}

\textbf{Energy and Area Consumption.}
For measuring the area and energy consumption of \thiswork, we utilized three distinct platforms for the graph processing accelerator, cache, and DRAM.
The energy consumption and area of the graph processing accelerator were measured by implementing it at the RTL level using Verilog HDL and synthesizing it with OpenROAD Flow~\cite{openroad_dac, openroad_gomactech}, which is an open-sourced RTL to GDSII tool.
We synthesized the accelerator up to placement and routing with a 45nm Nangate45 PDK (FreePDK45) library to run at 1GHz and scaled it to 22nm to match the tech node of the on-chip memory model.
We used the synthesis results, excluding SRAM.
To model the area, energy consumption, and latency of both the \ourcache design and a conventional cache, we utilized CACTI 7.0~\cite{cacti}.
We collect the energy per access through CACTI for the fg-tag array, which is similar to an 8-way set associative cache with an eight-bit tag. 
For the tag and data array, we collect the energy per access of the eight-way set associative cache using 128B cache lines.
Also, we estimate the energy consumption and access latency of \cq through CACTI in a similar manner.
By summing up the dynamic and leakage energy consumption of the tag array and the data array with \cq, we estimate the total energy consumption per access and latency of \ourcache.   
Similarly, for the on-chip cache area, we sum up the area of the two cache configurations above.
For DRAM die area overhead, prior academic conventions that synthesize modules with logic process nodes with scaling factors~\cite{gradpim, trim} are known to underestimate area overhead.
Instead, we performed a custom design and compared the modules to that of \cite{hynix}, which provides the area of each component of DRAM by reverse-engineering an existing product. 


\textbf{Baselines.}
To evaluate the performance of \thiswork, we compared \thiswork with five baselines.
First, we tested the conventional graph accelerating architecture described in graphicionado~\cite{graphicionado}.
Second, we tested the baseline accelerator architecture with scratchpad memory (SPM) described in~\cite{graphdyns}, and the accelerator interfaced with a conventional memory system alongside our baseline graph processing accelerator~\cite{graphdyns}, hereafter referred to as `GraphDyns (SPM)' and `GraphDyns (Cache)'. 
We set 4.5MB on-chip memory size to cache temporal vertex property ($V_{temp}$) for the above three baselines.
The size is slightly larger than that of \thiswork to compensate for its increased SRAM use in MSHR.
We also added a processing-in-memory (`PIM') solution to the baseline, which utilizes near-bank units that process the functions $Process, Reduce, Apply$ described in \cref{algo:vcpm} inside the off-chip memory similar to~\cite{graphpim}.
Lastly, we compared a near memory processing (`NMP') solution that implements the random scatter-gather in a buffer chip without adding extra area overhead in the DRAM chip like \cite{axdimm}.
Similar to \thiswork, NMP baseline can benefit from on-chip support because it just gathers/scatters the data from memory. 
For fair comparisons, all baselines employed graph tiling with the best tile width as determined by an exhaustive search.
Note that Graphicionado and GraphDyns (SPM) utilize graph tile width that perfectly matches the on-chip memory size for scratchpad memory.
Additionally, we compared our design to several fine-grain cache designs~\cite{amoeba, scrabble, graphfire}, applying slight modifications to each to get better performance for graph processing.

\begin{figure}[t]
    \centering
    \includegraphics[width=\columnwidth]{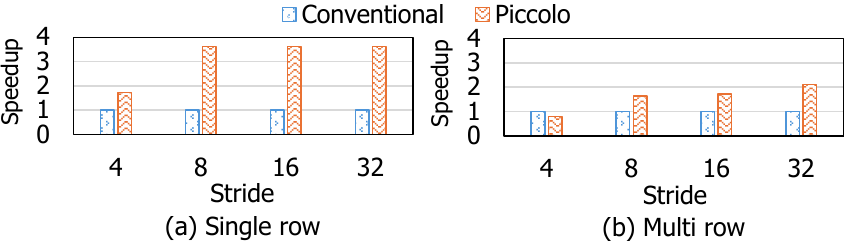}
    \vspace{-7mm}
    \caption{Microbenchmark result on the FPGA emulation. 
    Speedup is measured for reading 16MB data of varying strides.
    (a) Data are within a single row in each bank. (b) Data are distributed across multiple rows.}
    \vspace{-3mm}
    \label{fig:emul} 
\end{figure}

\begin{figure*}[ht]
    \centering
    \includegraphics[width=\textwidth]{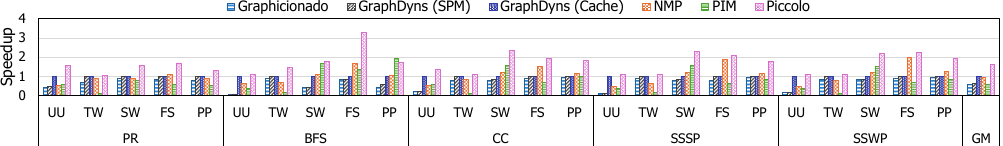}
    \caption{Speedup of \thiswork compared to various baselines.}
    \label{fig:perf} 
\end{figure*}

\begin{figure*}[ht]
    \centering
    \includegraphics[width=\textwidth]{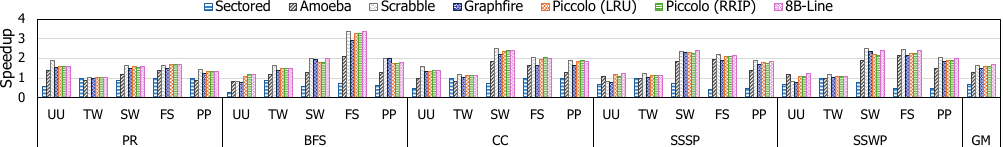}
    \caption{Effect of \ourcache and other fine-grained cache designs on top of \thiswork-FIM.}
    \label{fig:cachesense} 
\end{figure*}

\textbf{Graph Datasets and Algorithms.}
Table~\ref{tab:dataset} describes the graph datasets used for evaluations.
We chose five real-world graph datasets with various sizes and average degrees.
We tested PageRank~\cite{pagerank} (PR), Breath-First Search (BFS), Connected Component (CC), Single-Source Shortest Path (SSSP), and Single-Source Widest Path (SSWP) algorithms to evaluate \thiswork.
Each algorithm was conducted until convergence, with up to 40 iterations for cases where the number of iterations was too long. 
For the unweighted real-world graphs, integer weights between 0 and 255 were randomly assigned.
We additionally utilized synthetic graphs to evaluate the performance of the graph without a power-law distribution generated by the Watts-Strogatz model~\cite{watzstrogatz}, and to study scalability, we used Kronecker random graph generation~\cite{kronecker}.

\subsection{Microbenchmark via FPGA Emulation}
\label{sec:emul}

On the FPGA emulation platform, we verified the functionality of \thiswork-FIM and the commanding with DDR4 protocol as described in \cref{sec:main:impl}.
\cref{fig:emul} shows the measured results of a microbenchmark when \thiswork-FIM is applied on accesses with varying strides.
\cref{fig:emul}a depicts the performance comparison on data that fits into open rows of the banks.  
\thiswork-FIM achieves high speedup near the theoretical value of 4$\times$, which is reached at the stride of 8.
\cref{fig:emul}b shows the performance when the data are distributed across multiple rows. 
The speedup is relatively lower, as the activation latency takes a portion of the execution time. 
However, \thiswork still shows significant speedups, which will be higher with multi-rank systems that comprise more banks to hide the activation latency. 
One exception is the stride of 4, where the baseline penalty is halved because two elements fit within a 64B burst.

\subsection{Overall Performance}
\label{sec:eval:perf}

\cref{fig:perf} illustrates the performance comparison among five baseline methods and \thiswork. 
The final set of bars, labeled GM, represents the geometric mean across all evaluated algorithms. 
Among the baselines, we choose GraphDyns (Cache) and report normalized speedups because we modify the memory system based on GraphDyns (Cache), which performs the best among the baselines.
Overall, in geometric mean, \thiswork achieves a 1.62$\times$ speedup over the GraphDyns (Cache) and 1.68$\times$, 2.83$\times$ speedup compared to NMP and PIM baselines, respectively. 
Because Graphicionado and GraphDyns (SPM) utilize a scratchpad, which necessitates perfect tiling, they would not benefit from the graph sparsity.
For example, in the Uci-Uni (UU) dataset, whose average degree is three, those two baselines significantly underperform GraphDyns (Cache) because of the repetitive accesses to active vertices. 
The speedup of \thiswork over the baselines is mainly due to more efficient off-chip bandwidth utilization and fine-grained on-chip cache usage.
Unlike conventional systems, \thiswork-FIM can transfer data in fine granularity, storing only necessary data in \ourcache. 
On the other hand, the PIM baseline underperforms conventional methods because it cannot leverage the on-chip cache, leading to performance loss despite its internal memory bandwidth potential.
Although the NMP baseline (NMP) can utilize the on-chip memory support and the internal memory bandwidth at rank-level, it is far outperformed by \thiswork, which can utilize much higher bank-level internal bandwidth.

We also analyze the performance improvements across various graph algorithms and graph datasets. 
Over the baselines, \thiswork achieves larger speedups in active-vertex-based algorithms (BFS, CC, SSSP, and SSWP) that access only a subset of the edges each iteration. 
Unlike the PageRank (PR) algorithm, which accesses all edges in the graph during each iteration, those active-vertex-based algorithms access data more sparsely, resulting in a lower proportion of useful data from memory transactions.
In the Friendster (FS) dataset, \thiswork shows especially higher speedup because, as shown in \cref{fig:moti}, the conventional system shows a lot of unuseful portion both in accessed and cached data even in the perfect tiling case (more than around 80$\%$).
On the other hand, in the Twitter (TW) dataset, \thiswork provides a relatively lower speedup.
As vertices in TW are known to form dense clusters, they exhibit high-locality characteristics during processing. 
Therefore, \thiswork and the conventional systems on the TW dataset benefit substantially from on-chip memory support and high locality, which is also reflected in the notably lower performance observed in the PIM baseline.

\subsection{Comparison with Other Cache Designs}

\cref{fig:cachesense} shows the performance comparison among various alternative fine-grain cache designs: sectored cache~\cite{sectored}, amoeba-cache~\cite{amoeba}, scrabble cache~\cite{scrabble}, graphfire cache~\cite{graphfire}, \ourcache with different replacement policy (LRU and RRIP~\cite{rrip}), and 8B-line cache.
The performance is normalized to that of conventional 64B caches.
Due to the inefficient cache line usage as discussed in \cref{subsec:cachepolicy}, utilizing the sectored cache is significantly slower, which performs even worse than the conventional system.
Also, amoeba-cache and graphfire-cache achieve relatively lower performance because they store the metadata along with the cache data, resulting in lower effective cache capacity.
The 8B-line cache can store each fine-grained data from the memory independently, hence achieving the highest speedup against the others.
However, the overhead for cache line tags severely increases compared to 64B conventional cache lines.
From our careful designs, \ourcache performs almost like the 8B-line ideal case with much lower tag overhead as shown in \cref{sec:cache_arch}.
Compared to the 8B-line case, \ourcache exhibits only 3.90\% of the performance degradation in geometric mean.
Although scrabble-cache achieves similar speedup compared to 8B-line cache in geometric mean, their design complexity and metadata overhead are much larger than \ourcache design due to large additional metadata and comparators.
Additionally, another cache replacement policy, RRIP, could gain a marginal speedup.
However, it is insufficient to justify the additional overhead. 
This is because graphs exhibit random access patterns whose locality cannot be easily captured by general-purpose replacement policies.
Overall, \thiswork-cache achieves a good balance between cost and performance.


\subsection{Off-Chip Memory Access Analysis}

\begin{figure}[t]
    \centering
    \includegraphics[width=\columnwidth]{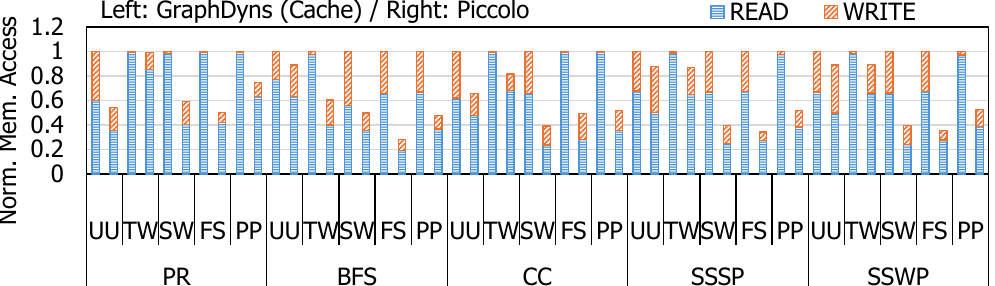}
    \caption{
    Normalized off-chip memory access breakdown (read and write) of the baseline and \thiswork.}
    \label{fig:memaccess} 
\end{figure}

\textbf{Memory Access. } 
\cref{fig:memaccess} shows the total off-chip memory transaction of \thiswork, normalized to that of the GraphDyns (Cache). 
Although our design newly introduces additional access to DRAM internal buffers (offset buffer, data buffer), \thiswork can scatter/gather random eight 8B data within a DRAM row by a single transaction. 
From this, \thiswork reduces total memory accesses by 43.2\% in geometric mean compared to the conventional system.
In the conventional system (GraphDyns), tiling support helps reduce the random access to the vertex properties by holding the data on the on-chip memory.
However, this increases the redundant access to the topology data, which increases read memory transactions.
On the PR algorithm, most of the datasets are found to be the fastest with perfect tiling, 
which slices a graph to small tiles that fit within on-chip memory.
Therefore, it significantly decreases the random access to vertex properties.
However, as shown in \cref{algo:vcpm}, an increased loop count from the perfect tiling requires a lot of additional memory read accesses.
In turn, the reduced memory transaction of \thiswork comes from two factors: 
allowing the use of larger tiles
and efficiently using off-chip bandwidth by \thiswork-FIM.

\begin{figure}[t]
    \centering
    \includegraphics[width=\columnwidth]{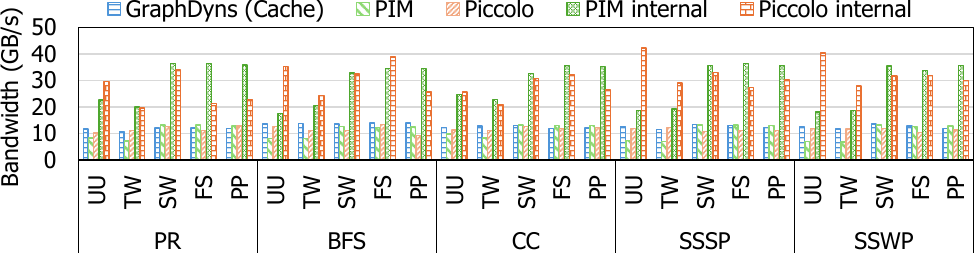}
    \caption{Off-chip DRAM and internal-DRAM bandwidth usage (GB/s) of the baselines and \thiswork.}
    \label{fig:bandwidth} 
\end{figure}

\textbf{Memory Bandwidth Utilization. } \cref{fig:bandwidth} illustrates the average bandwidth utilization of \thiswork and the two baselines. 
GraphDyns (Cache) and PIM utilize 65.5\% and 55.1\% the peak off-chip bandwidth utilization, respectively, and \thiswork attains 60.3\% utilization across five graph algorithms.
Compared to the baseline, \thiswork utilizes slightly lower off-chip bandwidth. 
However, this is due to the removal of unnecessary access, which is exemplified by the high internal bandwidth usage.
A similar pattern is found from the PIM baseline, but its performance is lower than \thiswork, as shown in \cref{fig:perf}, because of the lack of adequate usage of on-chip memories. 

\subsection{Energy and Area Analysis}
\label{sec:energy}

\begin{figure}[t]
    \centering
    \includegraphics[width=\columnwidth]{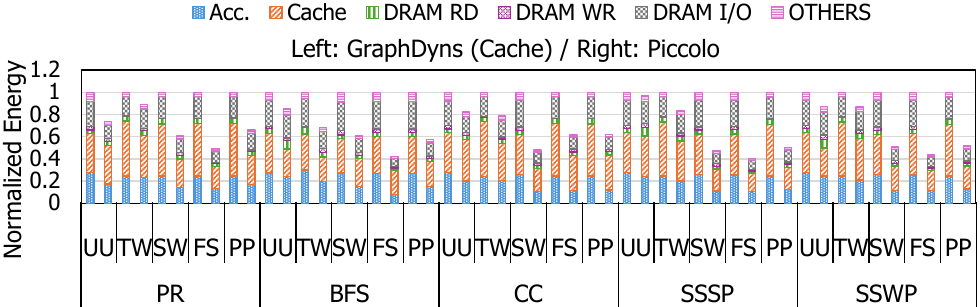}
    \caption{Energy consumption breakdown of \thiswork and the baseline with conventional DRAM.}
    \label{fig:energy} 
\end{figure}


\textbf{Energy Consumption.} \cref{fig:energy} shows the normalized energy consumption. 
The `Others' category represents the static energy of DRAM and refresh energy. 
Overall, \thiswork consumes 37.3$\%$ less energy in geometric mean compared to GraphDyns (Cache). 
The main source of energy saving is the reduction in the number of memory transactions.
As graph processing is highly memory-bound, memory accesses take a large portion of the energy consumption.
As shown in \cref{fig:memaccess}, \thiswork reduces memory transactions by 43.2\%, resulting in a substantial reduction in DRAM I/O energy, which is the largest portion of the DRAM energy consumption.
Because the amount of computation is equal, the energy saving from the accelerator mostly comes from the reduced static energy. 
While the cache energy consumption follows a similar pattern, it is also affected by larger tile sizes of \thiswork, which reduces the number of raw cache accesses. 
DRAM write energy slightly increases compared to the baseline because of \thiswork's requirement to write column offsets to the offset buffer for each scatter/gather operation.
However, this increment is negligible since the portion of the DRAM internal energy consumption is small.

\textbf{Area Overhead.}
We compare the area overhead of \thiswork compared to the conventional system.
The area of the accelerator, except for the on-chip cache, is calculated by RTL synthesis, and the area of the on-chip cache is calculated by CACTI. 
The final total chip area is 6.60mm$^2$, representing a 4.10\% increase over the conventional system’s area of 6.34mm$^2$. 
To estimate the overhead of \thiswork-FIM compared to conventional DRAM, we compare the internal controller and the offset/data buffers to the breakdown from \cite{hynix}.
\thiswork-FIM's internal controller includes a clock counter (4 counters, 72 transistors) to keep $tCCD\_L$, a command decoder (3$\times$2-bit AND, 18 transistors), and address offset buffer logic (6$\times$2-bit AND, 36 transistors), totaling 126 transistors. Compared to the 4096 transistors for CSL drivers and 2304 transistors for double-partitioned column decoders, this is significantly smaller, accounting for 0.04\% area.
For the offset and data buffers (128 bits per bank each), we conservatively assume the same per-bit size as local data buffers. According to \cite{hynix}, a 128-bit local data buffer accounts for 0.135\% of a 16Gb DDR4 die. Considering two additional buffers in each of the 16 banks, this totals 4.36\% overhead combined with the command generator.



\subsection{Sensitivity Studies}
\label{sec:sense}
We also conducted various sensitivity studies to validate the usability of \thiswork using all five graph algorithms on the Sinaweibo (SW) dataset.


\begin{figure}[t]
    \centering
    \includegraphics[width=\columnwidth]{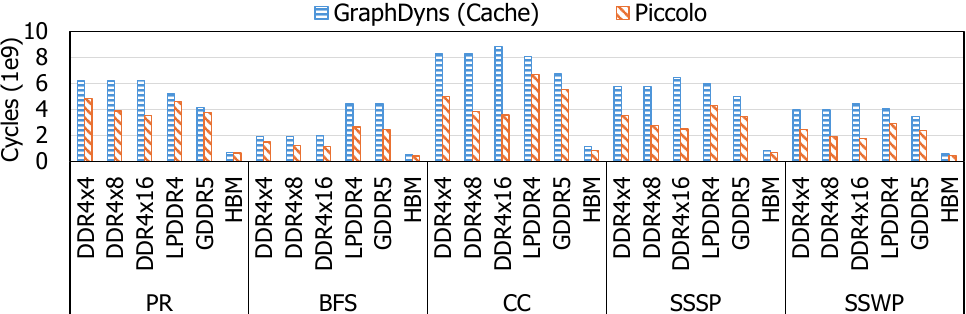}
    \caption{Memory type sensitivity of \thiswork.}
    \label{fig:memtype} 
\end{figure}

\textbf{Memory Type Sensitivity.} \cref{fig:memtype} presents the sensitivity to different memory types. 
Three DDR4 devices (x4/x8/x16), LPDDR4, GDDR5 and HBM devices are included. 
For DDR4 devices with smaller device widths, the number of devices becomes larger, which results in less speedup from more offset buffer write transactions. 
LPDDR, GDDR, and HBM have smaller burst granularity (32 bytes).
Therefore, there is less room for improvement from random scatter/gather than DDR devices because those scatter/gather four 8 bytes of data (a total of 32 bytes) in two memory transactions.


\begin{figure}[t]
    \centering
    \includegraphics[width=\columnwidth]{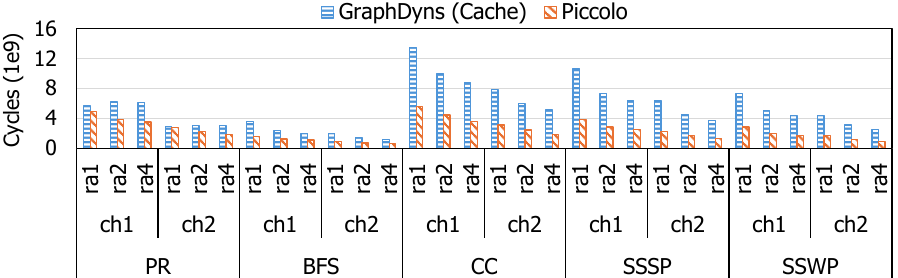}
    \caption{The number of DRAM channels and ranks sensitivity of \thiswork.}
    \label{fig:chra} 
\end{figure}

\textbf{Channel and Rank Sensitivity.} \cref{fig:chra} displays the sensitivity of \thiswork to the number of channels/ranks.
\thiswork provides more speedup since having more ranks indicates more banks. 
This means that \thiswork can enjoy higher internal bandwidth and a higher probability of hiding activation latency during scatter/gather.
For the active-vertex-based algorithms, although the trend is still maintained, the speedups compared to the baseline are similar across the number of ranks.
This is because the long latency induced by non-sequential accesses to topology leads to frequent row activations, thus the effect of hiding activation latency also highly affects the baseline.
Overall, \thiswork shows consistently better performance over GraphDyns (Cache) in different channel/rank configurations.

\begin{figure}[t]
    \centering
    \includegraphics[width=\columnwidth]{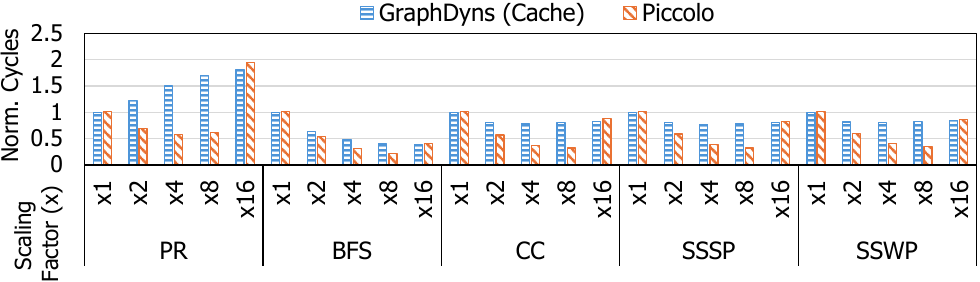}
    \caption{Tile configuration sensitivity of \thiswork.}
    \label{fig:tile} 
\end{figure}

\textbf{Tile Size Sensitivity.} \cref{fig:tile} shows the effect of the tile configuration. 
In \cref{fig:tile}, the leftmost bar of each graph algorithm represents the case of slicing the graph in a perfect tiling manner, which is denoted as $\times$1 (Scaling Factor 1).
The other bars, denoted as $\times$$n$ (Scaling Factor $n$), represent the cases where a tile size is $n$ times larger than the perfect tiling case.
Note that the best-performing tile sizes are different by graph algorithms.
On the PR algorithm, since it traverses all the edges, it can benefit more from locality.
Thus, it achieves the highest performance in perfect tiling, which houses all the required data in the on-chip cache.
On the other hand, \thiswork prefers larger tiles.
This is expected because \ourcache can hold only the useful data, which increases the effective cache capacity.
Furthermore, when the tile size exceeds the storage capacity of the on-chip cache, making it unable to accommodate all the necessary data, the randomness is more tolerable for \thiswork.
Likewise, the other algorithms show that \thiswork can tolerate the larger scaling factor and mostly perform better than GraphDyns (Cache).
However, when the tile size is set too large, \thiswork suffers performance degradation due to the frequent eviction of \cq.

\begin{figure}[t]
    \centering
    \includegraphics[width=.92\columnwidth]{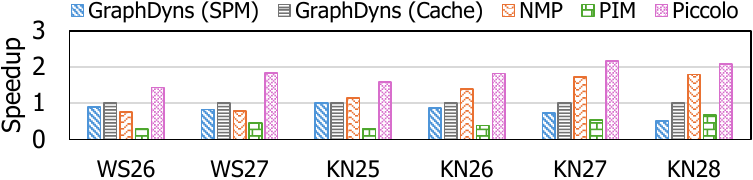}
    \caption{Speedup of \thiswork over baselines on various synthetic graphs.}
    \label{fig:synthetic} 
\end{figure}

\textbf{Synthetic Graph Analysis.}
We also evaluate the speedup using Watts-Strogatz~\cite{watzstrogatz} and Kronecker synthetic graphs~\cite{kronecker} for the PR algorithm across the four baselines and \thiswork.
As described in~\cref{fig:synthetic}, \thiswork outperforms the baselines for small-world networks (WS26, WS27).
This shows that \thiswork also works well for the graph datasets without power-law distribution.
Also, Kronecker synthetic graphs show the scalability.
GraphDyns (SPM) lacks scalability for large-scale graphs because of the overhead of tiling~\cite{graphdyns}.
PIM shows slightly better performance in larger graphs, but it still underperforms compared to GraphDyns (Cache), which benefits from the locality.
\thiswork consistently outperforms the baselines for all four graphs. 
As we mentioned, \thiswork-FIM architecture and \ourcache can benefit significantly from random accesses, making it scale well to larger graphs.

\begin{figure}[t]
    \centering
    \includegraphics[width=.91\columnwidth]{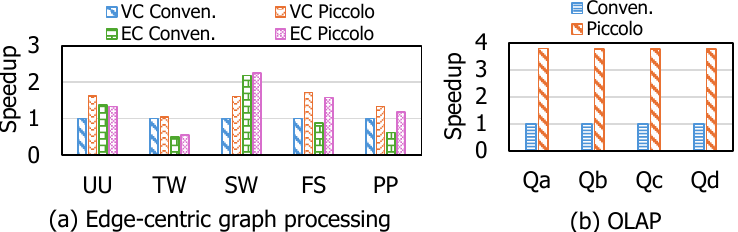}
    \caption{Comparison with the conventional setting in other environments. (a) in edge-centric graph processing accelerator. (b) in on-line analytic processing.}
    \label{fig:edge} 
\end{figure}

\subsection{Edge-Centric Graph Processing Model}
We further conduct experiments for edge-centric graph processing accelerators~\cite{foregraph, fabgraph, largescaleFPGA}.
Those accelerators also use the graph tiling approach to store all vertex properties for both source and destination nodes. 
Similar to~\cite{largescaleFPGA}, these random memory accesses to vertex properties create an opportunity for \thiswork-FIM and \ourcache to utilize memory bandwidth more efficiently. 
\cref{fig:edge}a presents the speedup normalized to a vertex-centric (VC) graph processing accelerator with a conventional memory system using the PR algorithm. 
\thiswork also achieves significant speedup for the edge-centric (EC) method, except for the UU dataset.
The average degree of the UU dataset is three, which is relatively low. 
Because of this, the graph's sparsity leaves less room for efficient scatter or gather for \thiswork.
Nonetheless, \thiswork with VC performs the best for the UU dataset.


\section{Discussion}

\subsection{Application on Other Domains}

\textbf{In-Memory Database.}
Even though we focused on graph processing, \thiswork could be beneficial for many workloads with fine-grained access patterns.
One good example is in-memory databases.
For online analytical processing (OLAP) queries scanning on specific columns, the individual data item are accessed with strides (usually 4 or 8 bytes). 
Similar to prior works~\cite{gsdram, RC-NVM, sam}, \thiswork will be beneficial to OLAP workload such as TPC-H benchmark.
To demonstrate this, we evaluate the four OLAP-style queries (depicted as Qa, Qb, Qc, Qd) from the RCNVMBench~\cite{RC-NVM} that comprise select statements, following prior work~\cite{sam}.
As shown in~\cref{fig:edge}b \thiswork-FIM can achieve about 3.8$\times$ speedup for OLAP queries compared to the conventional memory.

\textbf{Regular Applications.}
Aside from graph processing, the speedup of \thiswork could decrease for regular applications with good spatial locality.
For sequential access patterns, \thiswork-FIM would involve writing the unnecessary offsets, wasting the bandwidth.
To use \thiswork for general or mixed-purpose systems, we believe adding some locality monitor unit~\cite{spp, scrabble, amoeba} could alleviate the issue.
For example, when the system detects enough locality, the system can fall back to normal reads/writes.
When the pattern is mixed, separating the accesses into streams and applying cache partitioning with different methods would be an effective strategy.


\subsection{Design Choices}
\begin{figure}[t]
    \centering
    \includegraphics[width=\columnwidth]{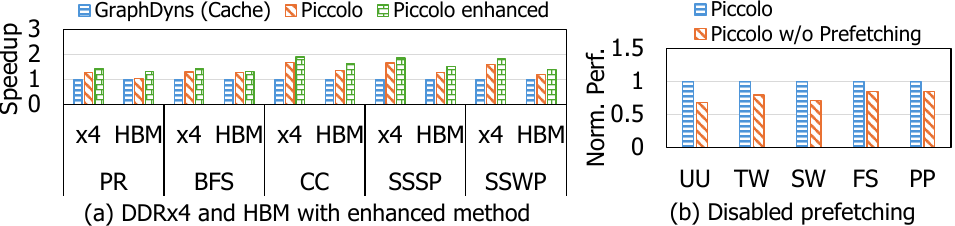}
    \caption{Comparison with baselines on other design choices. (a) on DDR4x4 with 11b column offset and HBM with supporting longer burst. (b) on a prefetching-disabled design.}
    \label{fig:design} 
\end{figure}

\textbf{Enhanced Design in Other Memory Types.}
As discussed in~\cref{sec:sense}, some memory types show less speedup than the DDR4$\times$16 case.
We can improve the performance for those cases with slight modifications to some design choices.
First, for devices with smaller widths (e.g., $\times$8 or $\times$4), we can use the column offset smaller than 16-bit because the typical DDR memory has a DRAM row size smaller than 8KB (lower than 11-bit per column).
This reduces the number of bursts needed for writing offset buffers.
Second, for memory types with a 32B burst, such as LPDDR, GDDR, and HBM, we can modify the memory chip to support a longer burst size.
Because a single burst is enough to provide eight offsets in those cases, we can expect more speedup.
\cref{fig:design}a shows the benefit of the above design choice. 
We evaluate the performance of DDR4$\times$4 device and HBM as representative examples of each case.
With the enhanced design, we achieve 17.9\% and 20.3\% additional speedup for each case in geometric mean. 

\textbf{Graph-Tailored Prefetching Designs.}
Many general purpose processors adopt graph-tailored prefetching designs to better utilize the cache~\cite{random1, graphfire}.
In graph processing accelerators, they also basically adopt the prefetching design to read topology data~\cite{graphicionado, graphdyns}.
To measure the effect of prefetching design, we compare the performance with the accelerator without prefetching.
\cref{fig:design}b shows the effect of disabling prefetching for the PR algorithm. 
Without prefetching design, \thiswork achieves 22.8\% slowdown in geometric mean.

\subsection{Comparison with Fine-Grained DRAM Architectures}
Many prior works~\cite{halfDRAM, fgDRAM, fgDRAM2, sectoredDRAM} try to use fine-grain row activation to reduce the energy waste, which stems from the coarse-grained activation granularity.
Half-DRAM~\cite{halfDRAM} splits the row into half, which enables fine-grained row activation. 
This has several benefits.
When the memory access patterns exhibit low locality, which incurs frequent row activation, half-DRAM can benefit from utilizing half-bank parallelism.
Also, the $tFAW$ constraint is critical for performance when issuing lots of activations.
By reducing the energy consumption with fine-grained activation granularity, the relaxation of $tFAW$ can boost the performance.
Similarly, Sectored-DRAM~\cite{sectoredDRAM} enables selectively activating each mat within a row and transferring only the selected words. 
Therefore, Sectored-DRAM can only activate and transfer useful data.

However, compared to \thiswork-FIM, those designs could not reduce the inefficient bandwidth usage.
Although Sectored-DRAM can transfer only useful data, the memory controller needs to wait for $tBURST$ even when the off-chip channel is idle.
Even when we assume shorter $tBURST$ to match the smaller transfer size, 
the problem with such a design is at the command bus overhead because a shorter burst does not reduce the address transferred to the memory. 
If a fine-grained burst is naively introduced, the current balance will sharply shift to the command bus, which becomes the new bottleneck~\cite{gradpim}.
Furthermore, \thiswork-FIM can utilize much higher internal bandwidth. 
Therefore, the benefit from \thiswork-FIM will be larger than other fine-grained DRAM designs.

\subsection{Limitations}
\label{sec:discussion}
\textbf{Flexibility.}
\thiswork performs random scatter/gather in an activated row. 
This raises some flexibility issues for scheduling when accesses do not fit well in a few rows.
However, in graph processing cases, they adopt the tiling approach to maximize cache hit, and the dynamic range of memory access can be significantly reduced.
Therefore, we can tune the graph tile size to be sufficient for graph processing.

\textbf{Overhead.}
In \cref{sec:main:impl}, we described how the \thiswork-FIM commands are implemented without adding a new command to DDR protocol.
However, if \thiswork were to be widely adopted, 
adding a dedicated command might be preferred.
We believe such a solution would not cause too much additional overhead. 
On the DRAM side, because the commands are already being executed, only a small change in the C/A encoding is necessary.
The memory controller needs to be modified to handle the new commands, whose overhead would be comparable to recently introduced commands such as masked write~\cite{lpddr5} or RFM~\cite{ddr5}.

\section{Related Work}

\subsection{Graph Processing Acceleration}
Graph processing~\cite{pagerank} is often described using diverse variations of programming models, and there are numerous approaches for accelerating it~\cite{graphmat,powergraph,tao,ligra,sisa}.
Some approaches use graph-tailored prefetching designs~\cite{random1, graphfire} targeting CPUs.
They put the prefetched graph data to different level caches depending on the access patterns of the data.
Some methods use GPUs~\cite{cusha,tigr, grassembler}, considering the imbalance~\cite{gunrock}, or reducing data transfer between GPU and CPU~\cite{subway,grus}. 
Graphicionado~\cite{graphicionado} and GraphDyns~\cite{graphdyns} uses highly parallel ASIC-based accelerator with vertex-centric programming model (VCPM)~\cite{pregel}.
They eliminate random memory access by using huge on-chip scratchpad memory to store all the vertex properties.
They utilize a tiling approach for large graphs that do not fit into the on-chip memory.
Fabgraph~\cite{fabgraph, fabgraph2}, Foregraph~\cite{foregraph} and MOMSes~\cite{stopcrying, largescaleFPGA} utilize FPGA-based methods using edge-centric programming model~\cite{xstream}.
Similar to the VCPM, they slice the graph into many blocks and process each block respectively.
Recently, some works utilize near-storage processing for graph processing~\cite{extrav} with a dedicated accelerator or accelerate graph neural network (GNN) inference/training with accelerators considering efficient on-chip memory usage~\cite{snf_cal,snf}, sparsity~\cite{sgcn} and distributed environments~\cite{granndis}.


\subsection{Mitigating Inefficient Memory Access}
Several approaches support variable cache line sizes~\cite{linedistil, amoeba, elasticcache, scrabble} to enable fine-grained data management, reducing the amount of unused data within the cache.
In addition, fine-grained DRAM architectures have been proposed by reducing activation granularity~\cite{halfDRAM, fgDRAM} and employing variable burst lengths to retrieve only necessary data~\cite{sectoredDRAM}.
Some approaches address the overfetching problem by supporting fixed-stride access patterns~\cite{gsdram, sam}.
On the other hand, to mitigate the memory bottleneck, prior works leverage the internal bandwidth through near memory processing~\cite{chameleon,fpganma,nda,axdimm,tensordimm,smartdimm} or processing in memory~\cite{truepim,newton,hbmpim,aim,gddraim}. 
Tesseract~\cite{tesseract} and GraphPIM~\cite{graphpim} utilize logic layers in 3D stacked memory~\cite{hmc} for graph processing. 
However, they often suffer from supporting various data types~\cite{upmem_benchmark,pidcomm,pidjoin,pathfinding}, which aligns with our point that arithmetic units in DRAM is expensive~\cite{hbmpim,aim,gddraim}.

\section{Conclusion}

Current graph processing accelerators utilize graph tiling or PIM approaches to address graph processing's fine-grained random access patterns.
However, those face significant limitations, especially when implemented on existing memory standards (i.e., DDR).
Therefore, we introduce \thiswork, an efficient end-to-end graph processing accelerator.
We introduce \thiswork-FIM, a function-in-memory (FIM) with in-memory random scatter-gather, and \thiswork-Cache, a redesigned cache and MHA to fully benefit from both the tiling-based and PIM approaches.
In extensive evaluations, \thiswork achieves 1.62$\times$ speedup and 37.3\% reduction in energy consumption in geometric mean.

\section*{Acknowledgements}
A preliminary work of \thiswork was published at IEEE CAL~\cite{piccolo_cal}.
Authors from SNU were supported in part by Institute of Information \& communications Technology Planning \& Evaluation (IITP) (2024-00395134, 
RS-2024-00347394) 
grant funded by the Korea government (MSIT) and in part by Samsung Electronics (IO230407-05813-01). 
FL was supported in part by US Department of Energy (DOE) Office of Advanced Scientific Computing Research (ASCR) under FWP ERKJ368.
Jinho Lee is the corresponding author.



\bibliographystyle{IEEEtranS}

\begin{flushleft}
\bibliography{refs}
\end{flushleft}

\end{document}